\documentclass[12pt]{article} 
\usepackage{psfig}
\usepackage{a4}

\newcommand{\khat}{{\bf \hat{k}}}
\newcommand{\phat}{{\bf \hat{p}}}
\newcommand{\vecE}{{\mathbf E}}
\newcommand{\veck}{{\mathbf k}}
\newcommand{\vecp}{{\mathbf p}}
\newcommand{\vecq}{{\mathbf q}}

\newcommand{\vecv}{{\mathbf v}}

\newcommand{\veckp}{{\mathbf k}'}
\newcommand{\bra}{\langle}
\newcommand{\ket}{\rangle}
\newcommand{\half}{\frac{1}{2}}

\newcommand{\pdotk}{\vecp\cdot\khat}
\newcommand{\omk}{\omega_{\veck}}
\newcommand{\omkp}{\omega_{\veckp}}
\newcommand{\ompk}{\omega_{\vecp+\veck}}

\newcommand{\omkmkp}{\omega_{\veck-\veckp}}

\newcommand{\omqmkp}{\omega_{\vecq-\veckp}}
\newcommand{\nk}{n(\omk)}
\newcommand{\npk}{n(\ompk)}
\newcommand{\nkcl}{n_{\rm cl}(\omk)}
\newcommand{\npkcl}{n_{\rm cl}(\ompk)}

\newcommand{\ncl}{n_{\rm cl}(k)}
\newcommand{\dncl}{n'_{\rm cl}(k)}
\newcommand{\ddncl}{n''_{\rm cl}(k)}

\newcommand{\ep}{\epsilon}

\newcommand{\lm}{\lambda}

\newcommand{\om}{\omega}  
\newcommand{\be}{\begin{equation}}
\newcommand{\ee}{\end{equation}}
\newcommand{\bea}{\begin{eqnarray}}
\newcommand{\eea}{\end{eqnarray}}
\newcommand{\bean}{\begin{eqnarray*}}
\newcommand{\eean}{\end{eqnarray*}}
\newcommand{\nn}{\nonumber}

\newcommand{\intk}{\int \frac{d^3k}{(2\pi)^3}}
\newcommand{\intkp}{\int \frac{d^3k^{\prime}}{(2\pi)^3}}

\begin{document}
\title{
\vskip -100pt
{  
\begin{normalsize}
\mbox{} \hfill \\
\mbox{} \hfill HD-THEP-99-49\\
\mbox{} \hfill ITFA-99-33\\
\mbox{} \hfill November 1999\\
\vskip  70pt
\end{normalsize}
}
Divergences in Real-Time Classical Field Theories 
at Non-Zero Temperature
}
\author{
Gert Aarts$^a$\thanks{email: aarts@thphys.uni-heidelberg.de}
\addtocounter{footnote}{1},
Bert-Jan Nauta$^b$\thanks{email: nauta@wins.uva.nl}
\addtocounter{footnote}{2} and Chris G.\ van Weert
$^b$\thanks{email: cvw@wins.uva.nl}
\addtocounter{footnote}{3}\\
\normalsize
{\em $\mbox{}^a$Institut f\"ur theoretische Physik, Universit\"at 
Heidelberg }\\
\normalsize
{\em  Philosophenweg 16, 69120 Heidelberg, Germany}\\
\normalsize
{\em $\mbox{}^b$Institute for Theoretical Physics, University of
Amsterdam}\\
\normalsize
{\em            Valckenierstraat 65, 1018 XE Amsterdam, the Netherlands}
\normalsize
}

\date{November 23, 1999}
\maketitle
   
\renewcommand{\abstractname}{\normalsize Abstract} 
\begin{abstract}
\normalsize 

The classical approximation provides a non-perturbative approach to
time-dependent problems in finite temperature field theory. We study the
divergences in hot classical field theory perturbatively.  At one-loop, we
show that the linear divergences are completely determined by the
classical equivalent of the hard thermal loops in hot quantum field
theories, and that logarithmic divergences are absent. To deal with
higher-loop diagrams, we present a general argument that the superficial
degree of divergence of classical vertex functions decreases by one with
each additional loop: one-loop contributions are superficially linearly
divergent, two-loop contributions are superficially logarithmically
divergent, and three- and higher-loop contributions are superficially
finite. We verify this for two-loop SU($N$) self-energy diagrams in
Feynman and Coulomb gauges.  We argue that hot, classical scalar field
theory may be completely renormalized by local (mass) counterterms, and
discuss renormalization of SU($N$) gauge theories.

\end{abstract}
 
\newpage
 
\renewcommand{\theequation}{\arabic{section}.\arabic{equation}}

\section{Introduction}
\setcounter{equation}{0}
\label{secintro}
The classical approximation \cite{Grigorev:1988bd} is a useful tool for
the study of infrared properties of quantum fields at high temperature
\cite{Smit:1997gn,Bodeker:1995pp,Arnold:1997dy,Arnold:1997yb,Aarts:1997qi,Aarts:1997kp,Buchmuller:1997yw,Nauta:1997yi}, 
which may be applied to calculate non-perturbative phenomena such as the 
Chern-Simons diffusion rate 
\cite{Ambjorn:1995xm,Moore:1997cr,Tang:1996qx}
(relevant for theories of baryogenesis 
\cite{Rubakov:1996vz}) and the dynamics of the electroweak phase
transition \cite{Moore:1997bn}, as well as real-time (plasmon)
properties of hot
non-abelian gauge theories \cite{Tang:1997gk}. The classical theory is
expected to be a good approximation
at low-energy because the classical limit $\hbar \rightarrow 0$ and the
low-energy limit of the Bose-Einstein distribution function $n$ yield the 
same result: 
\begin{equation}
\nk = \frac{1}{\exp (\beta \hbar \omk)-1}
\rightarrow 
\frac{1}{\beta \hbar \omk} \equiv \nkcl,\quad 
\hbar \omk  \ll T,  
\label{hbar0}
\end{equation}
where $\omk=\sqrt{\veck^2}$ is the frequency at wave-number $\veck$ and 
$\beta =1/T$ the inverse temperature.
Classical correlation functions are determined by a set of field equations in
Minkowski space and a thermal average over the initial fields at some
(arbitrary) initial time. 
In perturbation theory, classical vertex functions can be obtained by 
taking the limit $\hbar \rightarrow 0$ of the quantum expressions, which 
amounts to the replacement (\ref{hbar0}) of 
the Bose-Einstein distribution function by the classical 
distribution function. The resulting $\hbar$'s in the denominator are 
compensated by a positive power of $\hbar$'s arising from the loop 
counting of the diagrams under consideration, such that in the classical 
limit a non-trivial expression (and loops) remain.

The replacement (\ref{hbar0}) is a good approximation for infrared-dominated
diagrams, but it changes the ultraviolet behavior of the theory and 
introduces classical (Rayleigh-Jeans-type) divergences.  When the
classical theory is considered as a
low-energy effective theory, these divergences can be regularized by
introducing a cut-off of the order of the temperature, $\Lambda\sim
T/\hbar$. Since in a weakly coupled theory the temperature is large 
compared to dynamically generated energy scales such as $g^2T$, the 
resulting cut-off dependences are a direct reflection of the
divergences of the classical theory. The general strategy to improve the
effective theory is to include counterterms that reduce the cut-off
dependence. In particular, if a complete set of counterterms can be
specified, the cut-off may be sent to infinity and the theory is
renormalized.  It is clear that a knowledge of the divergences is
necessary to determine the appropriate counterterms. We will assume that
these divergences will be tractable in perturbation theory.

In the case of a $\lm\phi^4$ scalar field theory the divergences have
been studied in classical perturbation theory for the two-point function
up to two loops and the four-point function up to one-loop
\cite{Aarts:1997qi,Aarts:1997kp,Buchmuller:1997yw}. 
It was found that the
one-loop resp.\ two-loop correction to the self-energy is linearly resp.\
logarithmically divergent, and that the one-loop correction to the
four-point function is finite \cite{Aarts:1997kp}. In $3+1$ dimensional
gauge theories on the other hand,
the attention has mainly been restricted to the classical equivalent of
the quantum hard thermal loop (HTL) expressions
\cite{Braaten:1990mz,Taylor:1990ia,LeBellac:1996}, which introduce linear
divergences in the classical theory
\cite{Bodeker:1995pp,Arnold:1997dy,Arnold:1997yb}.  
Numerical studies using a HTL improved effective theory
\cite{Bodeker:1995pp,Hu:1997sf,Iancu:1998sg}
can be found in \cite{Moore:1998sn,Rajantie:1999mp,Bodeker:1999gx}.
An analysis of the divergences in the classical theory
that goes beyond the HTL limit at one-loop, or to higher loops, has not
yet been performed for gauge theories. Our aim in this paper is therefore 
to give a more complete analysis of the divergence structure of hot,
real-time classical field theory.

A different kind of  effectively classical theory beyond the HTL regime was
constructed in \cite{Bodeker:1998hm}, by integrating out the scales $T$
and $gT$ in a leading log approximation. It takes the form of a Langevin
equation and is free from ultraviolet divergences \cite{Arnold:1998cy}.
For a numerical implementation, see \cite{Moore:1998zk}.\footnote{This effective theory has been rederived with the help of
classical transport theory, using the concept of classical colored point
particles \cite{Litim:1999ns}. It should be clear that we study classical
fields instead of classical particles.}
However, our
focus in this paper is on classical Yang-Mills theory as it stands,
without any further integrating out to construct an effective theory.

We shall argue that both in SU($N$) gauge theory and in scalar field
theory with $\phi^3$ and $\phi^4$ interaction terms the divergences are
restricted to one- and two-loop (sub)diagrams. It will be shown that
classical one-loop diagrams that correspond to HTL's in the quantum theory
lead to linear divergences, while other one-loop diagrams are
finite in the classical theory. Also we present a general argument that
two-loop diagrams can at most give logarithmic divergences. This is
explicitly verified for two-loop self-energy corrections in SU($N$) and
scalar theories.

The paper is organized as follows. Classical one-loop diagrams are
analyzed in the next section, and diagrams with two loops and more in
section \ref{sectwoloop}.  In section \ref{secrenorm} the possibility of
absorbing the divergences with counterterms is discussed. The final
section contains the conclusions. Throughout the paper expressions for
classical diagrams are obtained by taking the $\hbar\to 0$ limit in the
quantum expressions.  The validity of this is discussed in more detail in
appendix \ref{apprules}, where a set of classical Feynman rules is
presented for hot scalar fields.

\section{One-loop}
\setcounter{equation}{0}
\label{seconeloop}

\subsection{Linear divergences: classical HTL's}
\label{seclinear}

The one-loop linear divergences of the classical theory are closely
related to the (quantum) hard thermal loops  discovered by Braaten and
Pisarski \cite{Braaten:1990mz} (see also 
\cite{Taylor:1990ia,Silin:1960,Weldon:1982aq}). 
For instance, the divergent part of the classical self-energy in SU($N$)
gauge theory can be obtained as the classical limit of the HTL self-energy
\cite{Bodeker:1995pp,Arnold:1997dy}. To be specific, the spatial part of
the retarded HTL self-energy reads\footnote{Loop momenta will generically
be denoted with $K=(k^0,\veck)$ and external momenta with $P=(p^0,\vecp)$. 
Furthermore, $K^2=-k_0^2+k^2$, $k=|\veck|=\omk$, and $\khat=\veck/k$.} 
\be 
\Pi_{HTL, ij}^{ab}(P)=-2\delta^{ab}g^2\hbar N \intk\,
\hat{k}_i\hat{k}_j n'(\omk) \frac{p^0}{p^0-\khat\cdot\vecp},
\label{spats-e} 
\ee 
where here and in the following the external frequency
$p^0$ is taken real with a small imaginary part to obtain the retarded
self-energy, i.e.\ $p^0 \equiv \mbox{Re}(p^0)+i\ep$, and 
\be
n'(\omk) = \frac{d\nk}{d\omk}.
\ee
As usual in the HTL approximation, the radial and angular integration
decouple and the radial integration determines the plasmon frequency 
\be 
\om_{\rm pl}^2=-\frac{1}{3\pi^2}g^2\hbar N
\int_0^\infty dk\, k^2 n'(k) = \frac{1}{9}g^2 N \frac{T^2}{\hbar}. 
\label{plasfreq} 
\ee

The classical self-energy corresponding to (\ref{spats-e}) is obtained by
taking the \(\hbar\rightarrow 0\) limit, before the integration over 
$\veck$ is performed. This simply amounts to replacing
the Bose-Einstein distribution function by the classical distribution
function, as in (\ref{hbar0}). The classical self-energy is non-vanishing,
since the \(\hbar\) in the prefactor of (\ref{spats-e}) is compensated by
the \(\hbar\) in the denominator of the classical distribution function. 
The resulting radial integral is linearly divergent and to handle this we
introduce a cut-off in the classical distribution function
\(\nkcl\rightarrow \nkcl\theta (\Lambda -k)\). This particular way of
introducing a momentum cut-off in loop integrals does not lead to problems
with gauge invariance, which can be most easily understood from the gauge
propagator of Landshoff and Rebhan 
\cite{Landshoff:1992ne} and is explained in appendix
\ref{appgaugecutoff}. The result is a linearly divergent classical 
plasmon frequency 
\be
\om_{\rm pl,cl}^2=\frac{2}{3\pi^2}g^2 N T \Lambda.
\label{clplasfreq}
\ee
The relation between the quantum plasmon frequency (\ref{plasfreq}) and
the classical analogue (\ref{clplasfreq}) is that the Bose-Einstein
distribution function effectively introduces a cut-off of the order of the
temperature on the integration, \(\Lambda\sim T/\hbar\).  Since the
angular integration is completely decoupled, the dependence on the
external momenta of the linearly divergent contribution to the classical 
self-energy and HTL self-energy are equal.\footnote{At least with a
(perturbative) continuum-like regularization as employed here. On a
spatial lattice, this is not the case
\cite{Bodeker:1995pp,Arnold:1997yb,Nauta:1999cm}.}
All of this is well-known \cite{Bodeker:1995pp,Arnold:1997dy}.

Hard thermal loops are the leading contributions to vertex functions for
soft external momenta $|p^0|,p\sim gT$. Power counting reveals that
one-loop diagrams, with any number of external gauge fields, contain a HTL
contribution. The fact that the external momenta are small compared to the
internal momentum $k\sim T$ allows for several simplifications in the
calculation of HTL's. As a result all HTL's are proportional to the
plasmon frequency squared (\ref{plasfreq}) 
\cite{Braaten:1990mz,LeBellac:1996}. 

Divergences in classical field theories have a similar behavior, since
here also the internal momenta $k\sim \Lambda$ are much larger than the
external momenta. In fact, all classical HTL's have the proportionality
factor (\ref{clplasfreq}). Therefore, all classical HTL's are linearly
divergent. 

Other one-loop contributions in the quantum theory are smaller by a factor 
\(p/k\sim p/T\). In the classical limit these subleading contributions
give a factor \(p/k\sim p/\Lambda\), which reduces the degree of divergence. 
Therefore we may conclude that all linear divergences at one-loop are 
given by the classical HTL's.

\subsection{No logarithmic divergences}
\label{seclog}

Next we will argue that there are no logarithmic divergences at one-loop
in the classical theory. Firstly, we discuss one particular
example in SU($N$) gauge theory explicitly, which is the spatial part of
the self-energy in the Feynman gauge. A convenient starting point is the
expression in the quantum theory, which
reads 
\bea
\nn
\Pi_{ij}^{ab}(P) &=& \delta^{ab}g^2\hbar N \intk 
\Bigg\{
g_{ij}\frac{2\nk+1}{\omk} - \frac{A_{ij}}{4\omk\ompk}\times\\
\nn
&&\bigg(
[\nk+\npk+1]
\Big[\frac{1}{p^0+\omk+\ompk}-\frac{1}{p^0-\omk-\ompk}\Big]\\
\nn
&&+[\nk-\npk]
\Big[\frac{1}{p^0-\omk+\ompk}-\frac{1}{p^0+\omk-\ompk}\Big]
\bigg)
\Bigg\},\\
\label{onels-e}
\eea
with
\be 
A_{ij}=\half\left[8k_ik_j+5p_ik_j+3k_ip_j+4(p^2-p_0^2)g_{ij}-2p_ip_j\right].
\ee
This diagram contains of course the HTL self-energy (\ref{spats-e}).
As before, the classical expression is obtained by taking $\hbar$ to zero. 
The non-thermal 
contribution from the ``1" in the first and second line vanishes as 
$\hbar$ goes to zero.

{}From the previous section we know that contributions to the self-energy
(\ref{onels-e}) are at most linearly divergent. The classical limit of the
momentum-independent tadpole-like contribution in the first line is indeed
linearly divergent. For the contribution proportional to $A_{ij}$, it
implies that the contributions bilinear in the external
momenta, i.e.\ the terms proportional to \(p_ip_j\) or
\(p^2\delta_{ij}\), can only give ultraviolet finite contributions, and
that the terms linear in the external momenta (terms proportional to
\(k_ip_j\) or \(p_i k_j\)) may give logarithmic divergences.  The
contributions proportional to \(k_i k_j\) may contain logarithmic
divergences besides the linearly divergent contributions as well.

To obtain the linearly and logarithmically divergent contributions we
expand the integrand in \(1/k\), so that we can estimate the ultraviolet
behavior of the integrand by power counting. The contribution from the
second line
reads 
\bea 
\frac{A_{ij}}{4\omk\ompk} [\nkcl+\npkcl]
\Big[\frac{1}{p^0+\omk+\ompk}-\frac{1}{p^0-\omk-\ompk}\Big] && \nn \\ 
=
\frac{A_{ij}}{4k^2} \left\{\frac{2}{k}\ncl+ (\vecp\cdot\khat)
\Big[\frac{1}{k}\dncl-\frac{3}{k^2}\ncl\Big] + {\cal O}(k^{-4})\right\}.&&
\label{expansion2} 
\eea 
The first term on the second line, with $A_{ij} \propto k_ik_j$, 
is  part of the
HTL contribution. The second term between curly brackets, 
and the first term with $A_{ij}\propto
p_ik_j, k_ip_j$, contain the contributions proportional to \(k^{-3}\), and
these may give a logarithmic divergence after integration. However, it
turns out
that these contributions are odd under the transformation
\(\khat\rightarrow-\khat\) and therefore they vanish upon integration. 
The other terms, including those indicated with ${\cal O}(k^{-4})$, are
ultraviolet finite by power counting.

Similarly, the third line can be expanded, and after some algebra it can
be written as
\bea
\frac{A_{ij}}{4\omk\ompk}[\nk-\npk]
\Big[\frac{1}{p^0-\omk+\ompk}-\frac{1}{p^0+\omk-\ompk}\Big]&& \nn \\
=
\frac{A_{ij}}{4k^2}
\frac{2\pdotk}{p_0^2-(\pdotk)^2}
\Bigg\{ (\pdotk)\dncl + \half(\pdotk)^2\ddncl&& \nn \\
-\frac{1}{k}\dncl
\bigg[(\pdotk)^2-p_0^2
\frac{p^2-(\pdotk)^2}{p_0^2-(\pdotk)^2}\bigg]
+{\cal O}(k^{-4})
\Bigg\}.&&
\label{expansion}
\eea
The first term on the second line, again with $A_{ij}\propto k_ik_j$, is
part of the HTL contribution, and is proportional to \(k^{-2}\). The other
terms
contain a contribution proportional to \(k^{-3}\), which after integration
could yield a logarithmic divergence. However, just as in the previous
case these contributions are odd under the transformation
\(\khat\rightarrow -\khat\) and they vanish upon integration. The
remaining terms are ultraviolet finite.

Therefore, we conclude that there is no logarithmic divergence in the
spatial part of the retarded classical self-energy in the Feynman gauge. 
In a similar manner, we have also verified that the spatial part of the 
three-point vertex contains no logarithmic divergences.

The reason for the vanishing of possible logarithmically divergent
contributions lies in the behavior of the self-energy and the vertex
functions under parity (P) and time reversal (T).  The spatial part of the
self-energy discussed here is invariant under $\vecp\to-\vecp$, and
$p^0\to -p^0$ in combination with complex conjugation (i.e.\ $p^0+i\ep \to
-(p^0+i\ep)$ in (\ref{onels-e})).  The point is that the expansion in
\(1/k\) turns out to be an expansion in PT odd (dimensionless) functions
of \(p^0\) and \(\vecp\). Since the linearly divergent HTL contributions
to the self-energy are even under P and T, the logarithmically divergent
contributions are odd and should therefore vanish.  This argument extends
to the temporal part of the self-energy as well as to other vertex
functions. 

Finally we would like to remark that the vanishing of logarithmic
divergences holds in general Coulomb or covariant gauges, since the 
corresponding gauge fixing term does not break PT invariance, and the
same argument can be applied.

\subsection{Classical self-energy: explicit result}
\label{secself-energy}

The analysis presented above is useful for a general understanding.
However, in some cases it is possible to actually calculate
the loop integrals 
and avoid an expansion in $1/k$. Here we give one of those explicit
results in SU($N$) theory.

We calculate the diagonal ($ii$) part of the classical one-loop retarded
self-energy in the Feynman gauge in appendix
\ref{appcalculation}, and the result reads
\be
\Pi_{ii,\rm cl}^{ab}(P) =
\delta^{ab}g^2N\left[ \frac{T\Lambda}{\pi^2}\frac{p^0}{p}\ln 
\frac{p^0+p}{p^0-p} +
\frac{T}{4\pi}\bigg( ip^0-\frac{3p^2-4p_0^2}{2p}i\ln 
\frac{p^0+p}{p^0-p}\bigg)\right].
\label{eqSUn}
\ee
The real and imaginary parts can be obtained in the usual way, using
\be
 \ln \frac{p^0+p}{p^0-p} = \ln \left|\frac{p^0+p}{p^0-p}\right| 
-i\pi\theta(p^2-p_0^2).
\ee

The linear divergence is precisely the equivalent of the hard
thermal loop contribution, which follows from the replacement
$T\Lambda/\pi^2\to T^2/(6\hbar)$.  The finite terms are exactly equal to the
terms linear in $T$ that are obtained in a high temperature expansion in
the quantum theory, as can be checked explicitly
\cite{Weldon:1982aq,Brandt:1991fs}.\footnote{Up to some typographic
errors.} There are no other terms. The $p^0\to 0$ limit equals the
well-known result from the quantum theory in the Feynman gauge
\cite{Kajantie:1985xx}
\be
\Pi_{ii,\rm cl}^{ab}(0,\vecp) =
-\delta^{ab}g^2N\frac{3pT}{8}.
\ee
Note that in this limit the leading-order (gauge-dependent) behavior is
completely determined by classical physics.

To conclude the one-loop analysis, the above described situation can be 
understood also directly by keeping $\hbar$ in the high-temperature 
expression of the quantum theory. 
The high-temperature expansion then 
has the form \cite{Weldon:1982aq,Brandt:1991fs}
\be
\Pi_{ii\rm}^{ab}(P) = \delta^{ab}g^2N\Bigg[
\frac{T^2}{\hbar}\Pi_{-1}(P) + 
T \Pi_{0}(P)+
\left(\hbar\ln \frac{T}{\hbar\mu}\right) \Pi_{\rm log}(P) + 
\hbar\Pi_{1}(P) + {\cal O}(\frac{\hbar^3}{T^2})\Bigg],
\ee
where $\mu$ is the renormalization scale. 
The term proportional to \(T^2\)
is the HTL part, which turns into the linearly divergent term when 
$\hbar\to 0$, and the second term in this expansion is the finite 
term in the classical theory. All the other terms vanish when $\hbar\to
0$.

\section{Two-loop and beyond}
\setcounter{equation}{0}
\label{sectwoloop}

\subsection{Degree of divergence}
\label{secdegree}

In this section we study the degree of divergence of higher-loop diagrams
in the classical theory.  In the first part we shall argue that the
superficial degree of divergence of the self-energy decreases
by one with each loop, starting with the one-loop linear divergence.
Then we will check this statement explicitly for a number of diagrams.
We shall argue that the same is true for classical vertex
functions in section \ref{sechigher}.
 
To make the argument for the self-energy, we start with the following
basic assumption: 
in the high-temperature limit the retarded self-energy in the quantum
theory scales according to its dimension, i.e., the quantum retarded
gluon self-energy behaves as 
\be
\Pi_{\mu\nu}(P)=T^2\bar{\Pi}_{\mu\nu}(p^0/p,\phat,g) +T^2{\cal O}(P/T),
\label{assumption}\ee
for high temperatures, fixed external momentum and frequency, and a
renormalization scale of 
the order of the temperature \(\mu\sim T\).
This assumption consists of two parts:
The contribution of diagrams with hard momenta \(K\sim T\)
on all internal lines gives a \(T^2\) contribution 
to the self-energy. 
Contributions that are 
excluded in (\ref{assumption}) are of the form 
\(g^{2L} T^2 (T/P)^{m}\) for \(m>0\) and with $L$ indicating the number of
loops. For {\em fixed} external momenta  and high temperatures such terms
become larger than the one-loop (HTL) contribution \(g^2T^2\), so they
invalidate a loop expansion. Therefore the assumed absence of these
contributions can be re-expressed by saying that we assume that hard modes
are perturbative.
The other part of the assumption is that also diagrams with soft internal
momenta give a \(T^2\) contribution. This relies on the belief that
infrared divergences are controlled by induced masses, which are 
proportional to the temperature, such as the electric and magnetic masses
in SU($N$) gauge theories.

Let us then consider a classical contribution to the self-energy
containing $M$ distribution functions. To be able to compare the degree of
divergence of such a contribution with the quantum expression, we regard
the temperature in the quantum self-energy as a particular ultraviolet
cut-off, and using the assumption (\ref{assumption}) we
count the degree of divergence as 2.  Since every classical
distribution function gives rise to an extra energy in the denominator
when compared to the quantum diagram,\footnote{In the ultraviolet regime
of a loop integral the quantum (Bose) distribution function can be
approximated as $\exp -\beta\hbar\om$ and acts as a cut-off function. On
the other hand, the classical distribution function remains proportional
to $1/\om$.} the classical contribution to the self-energy with $M$
distribution functions has then a superficial degree of divergence $2-M$.

To complete the argument, we now use that the number of
distribution functions $M$ can be related to the number of loops $L$ in
the following manner \cite{Guerin:1994ny,VanEijck:1995}. One way to obtain
the retarded self-energy is by using the imaginary-time or Matsubara
formalism \cite{Kapusta:1989,LeBellac:1996}. One first performs the sums
over the discrete loop frequencies
and then analytically continues the external frequencies to real values
with a
small positive part to incorporate the appropriate retarded boundary
conditions. In the imaginary-time formalism the
number of loops equals the number of Matsubara frequency summations. Using
the method of contour integration to perform these sums, each sum gives
rise to one `coth' function, either with positive or negative energy.
Explicitly, each sum gives a factor \cite{Guerin:1994ny,VanEijck:1995}
\be
\half \coth \frac{s\hbar\om}{2T} = n(s\om)+\half = s[n(\om)+\half],
\;\;\;\;s=\pm.
\ee  
Hence, the resulting expressions are of the form of spatial momentum
integrals over Bose-Einstein distribution functions, where the number of
distribution functions is equal to or less than the numbers of loops.  
The classical limit can now be taken by replacing 
$n(\om)+\half \to T/(\hbar\om)$,
such that the $\hbar$'s counting the loops cancel against the $1/\hbar$'s
from the distribution functions. After taking the classical limit, only
the leading term, which has as many distribution functions as loops,
remains and the number of classical distribution functions $M$ in a given
diagram is counted by the number of loops, $M=L$.  
Note that this applies not only to the self-energy diagrams but to 
vertex functions as well.
It follows then that the superficial degree of
divergence of a classical diagram is given by $2-L$, such that the
classical one-loop contribution to the self-energy is superficially
linearly divergent, the two-loop contribution is superficially
logarithmically divergent, and higher-loop contributions are superficially
finite.

\subsection{Two-loop self-energy diagrams} 
\label{secnontriv}

\begin{figure}
\centerline{\psfig{figure=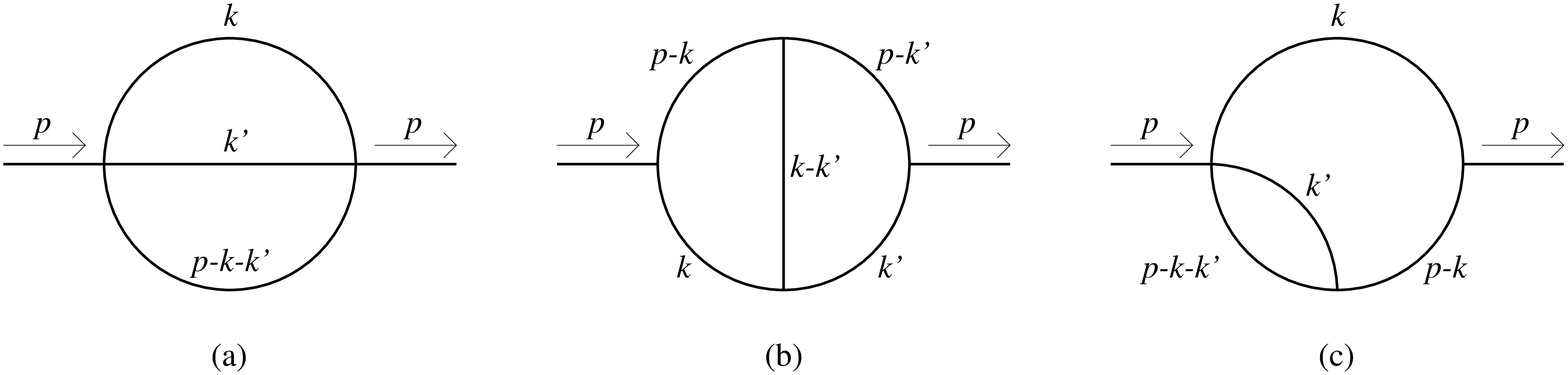,height=3cm}}
\caption{Two-loop diagrams. The setting sun diagram (a) and diagram (b)
are discussed in section \ref{secnontriv}, and diagram (c) 
is treated in appendix \ref{appnaive}.} 
\label{figtwoloopall} 
\end{figure}

We now want to verify the general argument of the previous
section for the two-loop self-energy diagrams appearing in SU($N$) and
scalar field theory.
We do not discuss diagrams which have a one-loop self-energy subdiagram
(and hence also a linear subdivergence), but we concentrate on the
two-loop diagrams as shown in fig.\ \ref{figtwoloopall}. 
Furthermore, since we are only interested in the structure of ultraviolet
divergences, i.e.\ in power counting, we do not need to make a
distinction between gauge field propagators in the Feynman gauge and
ghost propagators in the loops. 

Let us, as a first relatively simple example, take the two-loop
setting-sun contribution (a) to the retarded self-energy as it appears in
\(\lambda\phi^4\)-theory (with \(\lambda=g^2\)) and SU($N$) gauge theory.
It reads
\begin{eqnarray}
&&\Pi^{\rm (a)}(P)=  
\frac{1}{6}(g^2\hbar)^2
\intk\intkp \sum_{ss's_1}
\frac{ss's_1}{2^3\omk\om_{\veck'}\om_{\veck_1}}\,
\frac{1}{p^0+s\omk+s'\om_{\veck'}+s_1\om_{\veck_1}}  
\nonumber \\
&&
\Big\{
\left[1+n(s\omk)\right]
\left[1+n(s'\om_{\veck'})\right] 
\left[1+n(s_1\om_{\veck_1})\right]
-
n(s\omk) n(s'\om_{\veck'}) n(s_1\om_{\veck_1})
\Big\},  
\label{settingsun}
\end{eqnarray}
where $\om_{\veck_1} = \om_{\vecp-\veck-\veck'}$, and the sum is over all 
$s$'s being $\pm$. 

Note that the product of three distribution functions drops out. It is
then clear that the classical limit of (\ref{settingsun}),
\begin{eqnarray}
&&\Pi_{\rm cl}^{\rm (a)}(P) =
\frac{1}{6}g^4
\intk\intkp \sum_{ss's_1}
\frac{1}{2^3\omk\om_{\veck'}\om_{\veck_1}} \times \nonumber \\
&&
\hspace{1cm}
\frac{1}{p^0+s\omk+s'\om_{\veck'}+s_1\om_{\veck_1}}  
\left( 
s_1\frac{T^2}{\omk\om_{\veck'}} +
s'\frac{T^2}{\omk\om_{\veck_1}} +
s\frac{T^2}{\om_{\veck'}\om_{\veck_1}}
\right),  
\label{settingsuncl}
\end{eqnarray}
contains products of two classical distribution functions, 
in accordance with the statement that the number of loops
equals the number of distribution functions. 
We now estimate the degree of divergence by power counting and take the
loop momenta $k,k'\sim \Lambda$. The integral measures give two
contributions $\sim\Lambda^3$, and all single energy denominators $1/\om$
give a factor $1/\Lambda$.
The energy denominator that contains $p^0$ will produce, for generic large loop
momenta $\veck,\veck'$, a hard energy denominator $\sim 1/\Lambda$.
It can only produce a soft energy denominator when there is a
cancellation, which is in 
the special case that $\veck \simeq \pm\veck'$, depending on the signs of 
$s, s'$ and $s_1$ \cite{Braaten:1990mz}. However, for these
special configurations the integral over phase space is restricted so that
this will not alter the degree of divergence. We will use this estimate
for energy denominators with three hard energies \cite{Braaten:1990mz}
below as well.

By power counting we therefore establish that this contribution is
logarithmically divergent, as expected. This is also the result obtained
in \cite{Aarts:1997qi,Aarts:1997kp,Buchmuller:1997yw}, where the classical
setting sun diagram was analyzed in detail and it was shown that in fact
the logarithmic divergence can be separated and is independent of the
external momentum and frequency. 

It should be noted that the setting sun diagram (as well as the diagrams
discussed below) contains an infrared
divergence for vanishing external momentum \cite{Arnold:1994ps}. For
massless \(\lambda\phi^4\) theory, this can be cured by resumming the
effective thermal mass, arising from the one-loop tadpole diagram.
This has only an effect on the soft infrared modes, and does not interfere
with the ultraviolet behavior of the classical diagram we investigated
above. 

The next example we treat is the two-loop diagram (b) in 
fig.\ \ref{figtwoloopall}, which appears in SU($N$) and in scalar
$\phi^3$-theory. This particular diagram is more delicate, and it is
instructive to carry out the procedure described above in detail. 
We will verify explicitly that in SU($N$) theory (the spatial part of) 
this diagram is logarithmically divergent in the Feynman gauge. 

Since we are only interested in the degree of divergence of the diagram, 
we may ignore the color and Lorentz structure of the diagram. To indicate
the momentum-dependence of the four vertices in the gauge theory, we will
insert a factor $(k)^4_{ij}$. The precise form of the momentum insertions
is unimportant for the power counting.

We have found it convenient to calculate this diagram in the
imaginary-time formalism, and after performing the sums over the
Matsubara frequencies, the diagram can be written as
\begin{eqnarray} 
&& \Pi^{\rm (b)}_{ij}(P) = \half(g^2\hbar)^2 \intk\intkp\,
(k)^4_{ij} \sum_{ss's_1s_2s_3}
\frac{ss's_1s_2s_3}{2^5\om\om'\om_1\om_2\om_3} \times \nn \\ 
&&
\hspace{0.4cm}
\frac{1}{p^0+s'\om'+s_3\om_3}\; \frac{1}{p^0+s\om+s_2\om_2} \;
\bigg\{ \frac{1}{-s_3\om_3+s_2\om_2+s_1\om_1} \times \nn \\
&& 
\hspace{1cm}
\Big([n(s_1\om_1)+1][n(s_2\om_2)+1]n(s_3\om_3) - 
n(s_1\om_1)n(s_2\om_2)[n(s_3\om_3)+1] \Big) \nn \\ 
&& 
\hspace{0.4cm}
+\frac{1}{p^0+s_3\om_3+s\om-s_1\om_1} \times \nn \\
&&\hspace{1cm}
\Big([n(s_3\om_3)+1][n(s\om)+1]n(s_1\om_1) 
- n(s_3\om_3)n(s\om)[n(s_1\om_1)+1] \Big) \nn \\ 
&&
\hspace{0.4cm}
+\frac{1}{p^0+s'\om'+s_2\om_2-s_1\om_1} \times \nn \\
&&\hspace{1cm}
\Big([n(s'\om')+1][n(s_2\om_2)+1]n(s_1\om_1) 
- n(s'\om')n(s_2\om_2)[n(s_1\om_1)+1] \Big) \nn \\ 
&&
\hspace{0.4cm}
+\frac{1}{s'\om'-s\om+s_1\om_1} \times \nn \\
&&\hspace{1cm}
\Big([n(s_1\om_1)+1][n(s'\om')+1]n(s\om) - 
n(s_1\om_1)n(s'\om')[n(s\om)+1] \Big) 
\bigg\}, \nn\\
\label{twoloopex2}
\end{eqnarray}
where we have used the shorthand notation 
\be 
\om=\omk,\;\;\;\; \om'=\om_{\veck'},\;\;\;\; \om_1=\om_{\veck-\veck'},\;\;\;\;
\om_2=\om_{\vecp-\veck},\;\;\;\; \om_3=\om_{\vecp-\veck'},
\label{eqshorthand}
\ee
and the sum is over all sign factors $s=\pm 1$.  Again the products
of three distribution functions drop out.  The corresponding classical
integral $\Pi^{\rm (b)}_{ij,\rm cl}(P)$ may be obtained by taking the
\(\hbar\rightarrow 0\) limit, which amounts to neglecting the constants
and single distribution functions and replacing all
distribution functions that appear in products of two by classical
distribution functions. 

We will now consider the large $k,k'\sim\Lambda$ behavior of the classical
diagram as we did for the setting sun diagram, by looking at the various
factors in $\Pi^{\rm (b)}_{ij,\rm cl}(P)$ and naively combine those to
obtain an indication for its degree of divergence. 
First of all, each integration measure contributes $d^3k\sim \Lambda^3$
and the factor $(k)^4_{ij}$ is proportional to $\Lambda^4$. 
Each of the energies in the denominator on the first line gives a
contribution $1/\om \sim 1/\Lambda$, such that this factor leads to a
contribution $1/\Lambda^5$. Each classical distribution function gives
a factor $1/\Lambda$ as well.

The other energy denominators require a bit more care. 
All energy denominators between the curly brackets contain three large
energies that will generically not cancel, as in the case of the setting
sun diagram.  These therefore contribute with a factor $1/\Lambda$. 
The two energy denominators in the second line may produce a `soft'
energy denominator for specific combinations of the sign factors, namely 
for $s_3=-s'$, and $s_2=-s$. For example, the first denominator may give 
\be
\frac{1}{p^0 + s'(\om_{\veckp}-\om_{\vecp-\veckp})} \sim 
\frac{1}{p^0 + s'\khat\cdot\vecp} \sim\Lambda^{0},
\label{softden}\ee
similar to what happens in the one-loop case. This gives us four
possibilities: both energy denominators produce a soft contribution, only
one of them is soft and the other is hard, or both are hard.
Putting all these estimates together, we obtain in the first case, 
with two soft denominators, the naive result 
$\Pi^{\rm (b)}_{ij,\rm cl}(P) \sim
\Lambda^6\Lambda^4\Lambda^{-5}\Lambda^{-2}\Lambda^{-1}\sim\Lambda^2$,
which is a quadratic divergence. With one soft denominator we find
$\Pi^{\rm (b)}_{ij,\rm cl}(P) \sim \Lambda$, a linear divergence, and
with two hard contributions $\Pi^{\rm (b)}_{ij,\rm cl}(P) \sim
\Lambda^0$, the expected logarithmic behavior. However, from the general
argument we expect a logarithmic divergence only.

The reason for this mismatch is that this naive power counting
doesn't treat the distribution functions correctly.
In the one-loop (HTL) case, often differences of
statistical factors appear. In the classical theory, these lead to a
different ultraviolet behavior and hence change the power counting.
Therefore we take a closer look at the two-loop diagram to see whether a
similar thing occurs here as well. We denote the (naively) quadratically
divergent piece, with $s_3=-s'$ and $s_2=-s$, with
$\tilde{\Pi}^{\rm (b)}_{ij,\rm cl}(P)$. To re-estimate the divergence,
we put the external momentum in the energy denominator with three large
loop-energies (i.e.\ in the second, fourth and sixth line of
(\ref{twoloopex2})) equal to zero, since for generic large $\veck, \veckp$
the
denominator does not vanish.\footnote{Again, the region where it does
vanish is only a restricted part of phase space and is excluded in the
argument for power counting.} 
Taking the external momentum equal to zero can in fact be seen as the
zeroth order term in an expansion in the external momentum. The first
order term, linear in the external momentum, is treated in appendix
\ref{appnaive}.
The naively quadratically divergent contribution can now be written, after
flipping $s_1$ to $-s_1$ in the term on the sixth line, as
\begin{eqnarray} 
&& \tilde{\Pi}^{\rm (b)}_{ij,\rm cl}(P) = \half (g^2\hbar)^2
\intk \intkp \, (k)^4_{ij}
\sum_{ss's_1}
\frac{s_1}{2^5\om\om'\om_1\om_2\om_3} \, 
\frac{1}{p^0+s'(\om'-\om_3)}
\nonumber \\ 
&& 
\frac{1}{p^0+s(\om-\om_2)}\,\frac{1}{s'\om'-s\om+s_1\om_1}\, 
\left[n_{{\rm cl}}(s\om_2)-n_{{\rm cl}}(s\om)\right] 
\left[n_{{\rm cl}}(s'\om_3)-n_{{\rm cl}}(s'\om')\right].
\nonumber \\ && 
\label{naivequad} 
\end{eqnarray} 
We redo the power counting for $\tilde{\Pi}^{\rm (b)}_{ij,\rm cl}(P)$.
The thing to notice is that indeed two differences of two classical
distribution functions have appeared, and for hard loop-momenta
\begin{equation} 
\left[n_{{\rm cl}}(s\om_{\vecp-\veck})-n_{{\rm
cl}}(s\om_{\veck})\right] \sim -s(\khat\cdot\vecp) n_{\rm
cl}'(\omk)\sim\Lambda^{-2}. 
\label{distmindist}\end{equation} 
Both differences give one extra power of $1/\Lambda$, compared to the
naive
power counting employed before. The conclusion is therefore that
$\tilde{\Pi}^{\rm (b)}_{ij,\rm cl}(P)$, instead of being quadratically
divergent, is only superficially logarithmically divergent, as expected by
the general argument.

Note that the classical limit of diagram (b) may contain a linear
divergence from a HTL (three-point) subdiagram. The linear divergence
occurs, e.g.\ in contribution (\ref{naivequad}),  
whenever \((k)^4_{ij}\sim k^3k'\) or \((k)^4_{ij}\sim kk'^3\).
However, at this stage we are not interested in divergences caused by
one-loop subdiagrams since we study only the superficial degree of
divergence.

Potentially, there are also superficial linear divergences  in the classical
limit of (\ref{twoloopex2}).  
These are worked out in appendix
\ref{appnaive}. 
In this appendix we also discuss the other self-energy
contribution (c), which is naively linearly divergent as well. It turns
out that they are all in fact logarithmically divergent, in accordance
with the general argument of the preceding section.

\subsection{Higher-order vertex functions}
\label{sechigher}

We now extend the argument to general vertex functions. At
zero-temperature we know that the degree of divergence of a Feynman
diagram decreases with the number of external lines. In a real-time
classical theory at non-zero temperature this is not the case. We already
saw that the linear divergences at one-loop occur for diagrams with any
number of external gauge field lines.
Therefore we do not expect that the two-loop contributions to three- or
higher-point functions are finite.

To argue what happens for vertex functions with more loops, 
we use the real-time Feynman rules which are presented for
scalar field theory in appendix \ref{appscalar}. 
We employ Feynman rules in which two type of propagators appear, the
temperature-independent retarded propagator $G^R_0$ and the thermal
two-point function $S_0$ that contains the thermal distribution.
It is useful to recall here their explicit representation
\be
G_0^R(K) = \sum_{s=\pm} 
\frac{1}{2\omk}\frac{s}{k^0+i\ep + s\omk},\;\;\;\;
S_0(K) = 
\sum_{s=\pm} n_{\rm cl}(s\omk) \frac{2\pi s}{2\omk} 
\delta(k^0-s\omk).
\ee
Starting from the classical retarded
self-energy with $L$ loops (and hence $M=L$ thermal propagators),
generalized retarded $n$-point functions with $L$ loops can be obtained by
adding retarded Green functions in the loops, using the vertices (a) and
(c) shown in fig.\ \ref{figvertex} of appendix \ref{appscalar}. Note that
thermal propagators cannot
be added in the loops, since then the number of distribution functions $M$
is no longer equal to the number of loops, which is required by the
argument given in section \ref{secdegree} and is needed to have the
cancellation of $\hbar$ in the classical diagrams.  Note that this
also implies that all integrals over the zeroth components of
the loop momenta can trivially be performed with the help of the on-shell
delta functions in the thermal propagators.
To continue, in the case of a gauge theory, every additional
(momentum-dependent) 
three-point vertex gives an additional factor $K$ (we do not need to be
more specific for the power counting argument presented below).  Hence the 
total effect of adding one external line using a three-point vertex is an 
additional factor $K$ times a retarded propagator
\be
\label{eqGK}
 \frac{K}{\omk} \frac{s}{k^0+i\ep+s\omk}.
\ee
{}From the viewpoint of power counting, the first factor is of order
$\Lambda^0$, and the second factor can be of order $\Lambda^0$ or
$1/\Lambda$, depending on whether a soft or hard energy denominator
results, after the integrals over the on-shell delta functions in the
thermal propagators have been performed. 

\begin{figure}
\centerline{\psfig{figure=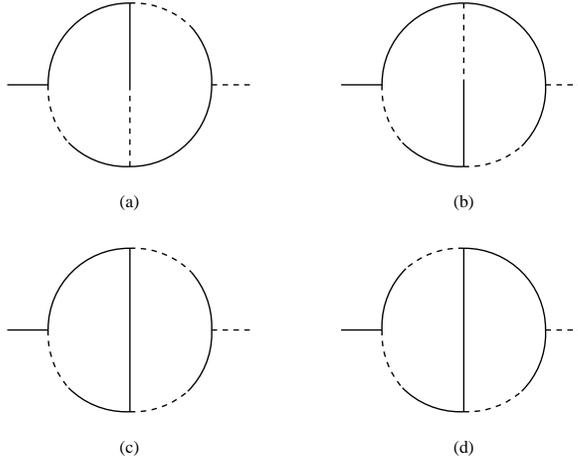,height=6cm}}
\caption{Two-loop diagrams in the real-time formulation that contribute
in the classical limit. Full lines
are thermal propagators and dashed-full lines retarded propagators.} 
\label{figtwolooprealtime} 
\end{figure}

This leads us to give the following general argument: in the case that 
the propagator in (\ref{eqGK}) is soft, the additional external line will not
change the degree of divergence, compared to the diagram without the
additional line. On the other hand, when the energy denominator turns out
to be hard, when the extra vertex is a 4-point vertex, or in scalar
field theory, where the  momentum $K$ in the numerator is absent,
additional lines will always lower the degree of divergence. Using the
result for the two-point function, this implies that higher-point vertex
functions are superficially logarithmically divergent by power counting
(at two-loop) or finite (at higher-loop).

There is one slight complication in this general argument.  In the
self-energy considered in the previous section, the logarithmic divergence
was the result of a subtle cancellation between quadratically (and
linearly) divergent contributions.  The question is whether this subtle
cancellation is not spoiled by adding an external line.  
Although a complete analysis of two-loop vertex functions is beyond the
scope of this paper, we will check explicitly in one particular case that
the cancellation indeed still occurs.

This analysis can be done most conveniently using the real-time
Feynman rules of appendix \ref{apprules}. We start by presenting in fig.\
\ref{figtwolooprealtime} the classical two-loop contribution to the
self-energy (b) in the real-time formalism. The integral over the zeroth
components of the loop momenta can easily be performed using the
on-shell delta functions in the thermal propagators, and we have verified
that this yields indeed the classical limit of (\ref{twoloopex2}), which
was calculated in the imaginary-time formalism, as expected.

\begin{figure}
\centerline{\psfig{figure=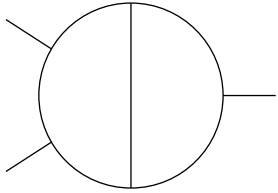,height=2.5cm}}
\caption{Two-loop diagram with three external lines.}
\label{figtwoloop3ext} 
\end{figure}

\begin{figure}
\centerline{\psfig{figure=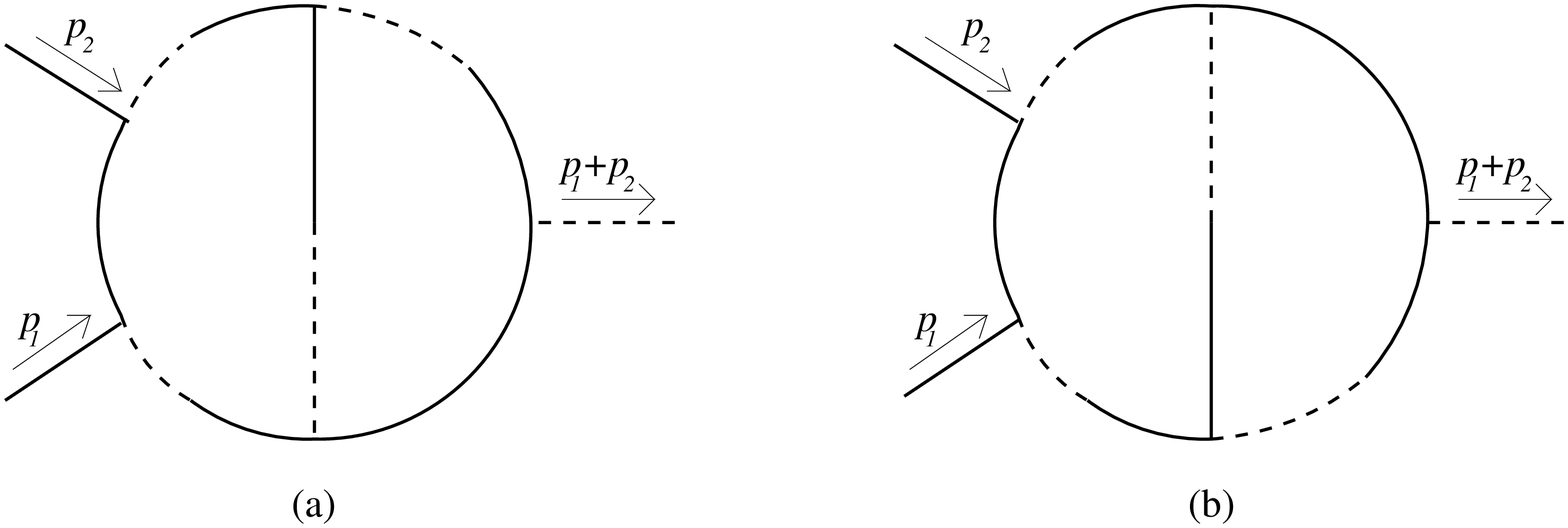,height=3cm}}
\caption{Two-loop contributions to the classical 3-point vertex function
in the real-time formalism that combined yield a logarithmic degree of
divergence.}
\label{figtwoloop3} 
\end{figure}

We want to add one external line to obtain a diagram as in fig.\
\ref{figtwoloop3ext}. 
In the case of the self-energy that we discussed in the previous section
we found that the naively quadratically divergent contribution
(\ref{naivequad})  does not contain a distribution function at energy
\(\om_1=\omkmkp\). That means that in terms of the real-time diagrams no
diagram with a thermal propagator on the line shared by the two loops
contributes to (\ref{naivequad}). Hence we do not need to consider the
addition of extra lines to the third and fourth diagram. Let's now see how
an additional three-point vertex of type (a) in fig.\ \ref{figvertex} 
and an additional retarded
Green function can be added to the first two diagrams in fig.\
\ref{figtwolooprealtime}. It turns out that
for each diagram (a) and (b) there are 14 possibilities to do this.  A
closer look, however, reveals that not all diagrams are needed to
establish a cancellation of the naive quadratic divergence.  For example,
a combination of the two diagrams that are shown in fig.\
\ref{figtwoloop3} is sufficient to obtain a difference between
distribution functions that reduces the degree of divergence to a
logarithmic one. 

Indeed, the sum of the most divergent part of the diagrams in fig.\ 
\ref{figtwoloop3} yields
\bea
\Gamma^{\rm (a+b)}_{ijk, {\rm cl}} 
&=& g^5\hbar^2\intk \intkp \, (k)^5_{ijk} \sum_{ss's_1} 
\frac{ss_1}{2^6\omk^3\omkp^2\omkmkp}
\nn\\
&&
\frac{1}{p^0_1+s\vecp_1\cdot \khat}\,
\frac{1}{p^0_2+s\vecp_2\cdot \khat}\,
\frac{1}{q^0-s'\vecq\cdot \khat^{\prime}}\,
\frac{1}{s_1\omkmkp-s\omk-s'\omkp}
\nn\\
&&
\left[n_{\rm cl}(s\om_{\vecp_1-\veck})-n_{\rm cl}(s\omk)\right]
\left[n_{\rm cl}(s'\omqmkp)-n_{\rm cl}(s'\omkp)\right],
\eea
with \(Q=P_1+P_2\). The factor \((k)^5_{ijk} \) has been included to
account for the momentum insertions from the vertices in a SU($N$) gauge
theory, and the factor $\hbar^2$ arises from loop-counting. We had to
expand also the single energy denominators, such as
\(1/\om_{\vecp_1-\veck}\), in external
momenta. Compared to the self-energy expression (\ref{naivequad}) 
the vertex function
has one extra factor (\ref{eqGK}) with a soft energy denominator as 
anticipated.
After power counting, taking into account (\ref{distmindist}), we
may conclude that in this particular combination the addition of one
external line does not spoil the reduction from a quadratic divergence to
a logarithmic divergence.

It will be interesting to make explicit checks for other three- (and
higher) point vertex functions with two loops as well, but without a
clever method to combine the different contributions this seems to be out
of the question.

\subsection{Other gauges}
To verify the general argument in section \ref{secdegree} that 
two-loop diagrams are logarithmically divergent, we have 
estimated in sections \ref{secnontriv} and \ref{sechigher}
the degree of divergence of some two-loop diagrams in the Feynman gauge.
Here we want to argue that the estimates in the Feynman gauge extend to 
general Coulomb gauges \cite{Braaten:1990mz}.

The gauge propagator in a general Coulomb gauge with gauge parameter 
\(\alpha_C\) reads
\be
\Delta_{\mu\nu}=\frac{1}{K^2}T_{\mu\nu}(\veck)+
\delta_{\mu 0}\delta_{\nu 0}\frac{1}{k^2}+\alpha_{C}\frac{K_{\mu}K_{\nu}}{k^4},
\ee
with the transverse projector 
\(T_{ij}(\veck)=\delta_{ij}-k_i k_j/k^2, T_{00}=T_{0i}=T_{i0}=0\).

First we realize that the external momentum dependence in the transverse
projector may be neglected \(T_{\mu\nu}(\vecp-\veck)\sim
T_{\mu\nu}(\veck)\)  when the integration momentum \(\veck\) is large. In
the power counting of a diagram we may estimate \(T_{\mu\nu}\sim 1\), and
we see that a diagram with all transverse propagators has the same degree
of divergence as the same diagram in the Feynman gauge.  Since the
\(00\)-component and the gauge dependent part of the propagator cannot
give a soft denominator like (\ref{softden}), we can also neglect the
external momenta in these components, they are then estimated as
\(k^{-2}\). Therefore diagrams containing these components of the
propagator will not have a larger degree of divergence. We conclude that
the degree of divergence of a certain diagram is the same in the Feynman
gauge and in a general Coulomb gauge. We stress that this does not
necessarily imply that the logarithmically divergent contribution is gauge
independent as is the case for the linear divergences, this remains a
subject for further study.

Finally we like to remark that in general covariant gauges it is not 
expected that individual diagrams obey the power counting of section 
\ref{secdegree}, but rather the sum of the diagrams with a certain number of 
loops. 

\section{Renormalization}
\setcounter{equation}{0}
\label{secrenorm}

\subsection{Scalar field theory}

The general analysis given in the previous sections yields the following 
result for a scalar field theory with 
\be
V_{\rm int} = \int d^3x \left(\frac{1}{3!}g\phi^3 + 
\frac{1}{4!}\lm\phi^4\right)
\ee
as interaction term:
time-dependent classical scalar field at finite temperature is 
renormalizable.

To see this in more detail, let's start at one loop: first of all, the
tadpole diagram, which is the only HTL in the quantum theory, is linearly
divergent, as expected. But since this divergence is trivially independent
of the external momentum, it can be canceled with a mass counterterm
\cite{Bodeker:1995pp,Aarts:1997qi}. The self-energy correction with two
$\phi^3$ vertices is finite. This can be seen in a number of ways: the
general analysis revealed that in gauge theories such a diagram is
linearly divergent. In a scalar theory, the vertices are momentum
independent and this brings down the superficial degree of divergence with
2, which makes the diagram is finite. In other words, it is not a HTL. It
follows also from a explicit calculation \cite{Aarts:1997kp}, where the
one-loop correction to the vertex function in a
$\lm\phi^4$-theory was studied, 
which is equal to the one-loop self-energy we are
discussing now. It was found there that the classical result is the
leading order term in a high temperature expansion in the quantum theory.  
Finally, in the quantum theory one-loop contributions to higher $n$-point
functions are no HTL's either, therefore these diagrams will be
ultraviolet finite in the classical limit. The reason is that the extra
retarded Green functions that appear in the loop bring down the
superficial degree of divergence with at least one, and this cannot be
compensated by momentum-dependent vertices as in the gauge theory.

At two loop, the setting sun contribution to the self-energy is
logarithmically divergent as expected from the general analysis. This has
been calculated explicitly in
\cite{Aarts:1997qi,Aarts:1997kp,Buchmuller:1997yw}.  However, it has been
shown there that the divergent part can be isolated and is independent of
the external momentum and frequency: again the divergence can be absorbed
with a mass counterterm \cite{Aarts:1997qi}. For the other contributions
to the self-energy that were discussed in detail in the previous section,
we note the following: in the gauge theory, diagram (b) contains a power
counting factor $(k)^4\sim \Lambda^4$ due to the momentum-dependent
vertices. In the scalar case this is of course absent.  Hence the naive
quadratically divergent contribution in the gauge theory is superficially
finite in the scalar case, even without any need for subtle cancellations.
The same is true for diagram (c), here a power counting $(k)^2\sim
\Lambda^2$ is absent and the classical diagram is immediately
superficially finite as well.  We conclude that for the two-loop
self-energy corrections (neglecting diagrams that contain self-energy
subdiagrams) the setting sun diagram is the only diagram that is
superficially divergent. The absence of momentum-dependent vertices also
simplifies the analysis of $n$-point vertex functions. It was argued in
section \ref{sechigher} that the superficial degree of divergence goes
down at least by one if more retarded Green functions are added in the
loops in a scalar theory. This can be immediately applied here. The
complication in gauge theories, i.e.\ the requirement of a subtle
cancellation, is not needed for the vertex functions either, because even
without the cancellations, the diagrams are superficially finite.

Since the general analysis already shows that three loops and higher are
finite, we arrive at the conclusion formulated at the beginning of this
section: classical scalar field theory at finite temperature can be
renormalized with merely a mass renormalization. 
For more details on this and on e.g.\ the choice of finite parts, see
\cite{Aarts:1997kp}.

We would like to stress that the analysis has been limited to simple
correlation functions, as discussed above. Other quantities, such as
transport coefficients like the viscosity \cite{Jakovac:1998jd}, which
require expectation values of composite operators \cite{Jeon:1995if}, are
not covered here.

\subsection{One-loop renormalization in SU($N$) theory}

In section \ref{seconeloop} 
we have seen that the divergences at one-loop are given
by the classical HTL's. Below we will present counterterms for these
one-loop divergences. 

Since we are interested in a classical 
theory it is appropriate to consider the equations of motion
\be
\delta_{A_{\mu}}S=\left[D_{\nu},F^{\nu\mu}\right]=
\delta j^{\mu},
\label{lineqmot}\ee
with the covariant derivative \(D_{\nu}=\partial_{\nu}+igA_{\nu}\), and
the field strength
\(F^{\nu\mu}=\partial^{\nu}A^{\mu}-\partial^{\mu}A^{\nu}
+ig[A^{\mu},A^{\mu}]\).
The source \(\delta j^{\mu}\) consist of a finite renormalization and 
a divergent part. We write
\be
\delta j^{\mu}=j^{\mu}_{\rm fin}-j^{\mu}_{\rm div}.
\ee
The divergent source \(j^{\mu}_{\rm div}\) has to be chosen such that it
cancels the linear divergences \cite{Bodeker:1995pp}. 
The identification of the linear
divergences as classical HTL's then implies that 
the divergent source must equal the classical HTL induced source: 
\(j^{\mu}_{\rm div}=j^{\mu}_{HTL, {\rm cl}}\).
We take the finite renormalization such that the 
classical theory (\ref{lineqmot}) is consistent with 
the quantum results. 
This can be seen as the real-time analog of matching employed in
static dimensional reduction and the construction of effective
3-dimensional theories \cite{Kajantie:1996dw}.
Therefore, the finite renormalization should equal the quantum HTL's:
\(j^{\mu}_{\rm fin}=j^{\mu}_{HTL}\).
In a (classical) perturbative expansion this renormalization prescription 
means that any classical HTL (sub)diagram is replaced by the corresponding
quantum HTL diagram.

The HTL effective action is non-local, but the introduction of an
auxiliary field allows for a local formulation of the HTL equations of
motion \cite{Blaizot:1994be}.
The sources then read 
\bea
&&j^{\mu}_{\rm fin} = j^{\mu}_{HTL}=
2g^2\hbar N\intk\, n'(\omk) v^{\mu} W(x,\vecv),
\nn\\
&&j^{\mu}_{\rm div} = j^{\mu}_{HTL, {\rm cl}}=
2g^2\hbar N\intk\,n'_{\rm cl}(\omk) v^{\mu} W(x,\vecv),
\label{sources}
\eea
with the velocity \(v^{\mu}=(1,\vecv)\) and \(\vecv=\khat\).
The velocity-dependent auxiliary field \(W(x,\vecv)\) 
satisfies the equation
\be
\left[v_{\mu}D^{\mu},W(x,\vecv)\right]=\vecv\cdot\vecE(x),
\label{eqW}
\ee
with \(\vecE\) the (chromo-)electric field.

{}From (\ref{sources}) we see that the two sources can be combined into one 
expression
\be
\delta j^{\mu}=2g^2\hbar N\intk\,
[n'(\omk)-n'_{\rm cl}(\omk)] v^{\mu} W(x,\vecv).
\ee
As usual in the HTL approximation the radial integration can be performed
independently, 
which yields
\be
\delta j^{\mu}=3\delta\om^2\int\frac{d\Omega}{4\pi}v^{\mu}W(x,\vecv),
\label{comb}\ee
with the linearly divergent coefficient 
\cite{Aarts:1997ve,Iancu:1998sg}
\be
\delta\om^2=\om_{\rm pl}^2-\om_{\rm pl, cl}^2
\label{coef}\ee
(see section \ref{seclinear}).
We see that the whole set of non-local linear divergent
vertex functions
can be renormalized by the adjustment of one parameter. This is a special
feature of the continuum HTL's. 
On a lattice the counterterms are more
complicated \cite{Bodeker:1995pp,Arnold:1997yb,Nauta:1999cm}, since the
lattice breaks rotational invariance and the velocity on the lattice is
not equal to the speed of light. The effective theory with continuum-like
counterterms (\ref{comb},\ref{coef}) has been studied numerically in
\cite{Bodeker:1999gx}.
An important observation is that the equations
(\ref{lineqmot}, \ref{eqW}, \ref{comb}, \ref{coef}) have conserved energy
\cite{Blaizot:1994be}, such that a thermal average over initial fields can 
be defined and the equations are a part of a proper
classical statistical theory.
Furthermore, it is well-known that in the classical theory only the
combination $g^2T$ appears and $\hbar$ is absent. However, since $\om_{\rm
pl}^2 \propto g^2T^2/\hbar$, $T/\hbar$ is introduced as an independent,
non-trivial scale in the effective theory described above.

That no linear divergences occur in retarded vertex functions calculated
with the classical theory (\ref{lineqmot}, \ref{eqW}, \ref{comb}, 
\ref{coef}), can be seen by introducing a background field and
integrating
out the classical fields. This generates in the HTL approximation the
classical induced source which is precisely what is subtracted on the
r.h.s.\ of (\ref{lineqmot}). It would be nice to see in a perturbative
calculation that these counterterms are sufficient to absorb the linear
(sub)divergences in (superficial logarithmically divergent) two-loop
diagrams, but this has not been attempted here.

\section{Conclusion}
\setcounter{equation}{0}
\label{secconclusion}

Classical thermal field theories contain ultraviolet divergences. In an
analysis of classical vertex functions, we found that at one-loop
only linear divergences occur, which come from classical HTL's, i.e.\ the
classical equivalences of the HTL's in the quantum theory. Furthermore we
argued that for $n$-point vertex functions with arbitrary $n$, the degree
of divergence decreases with the number of loops. This implies that
two-loop contributions are (superficially) logarithmically divergent and
higher loops are superficially finite. This may be compared with static
dimensional reduction, where the \(L\)-loop contribution to the
self-energy has also a degree of divergence \(2-L\). The difference is
that in the static limit higher-point vertex functions are less divergent 
than the self-energy. 
Indeed, the static theory is a superrenormalizable field theory and
a finite number of counterterms, like a one- and two-loop mass
counterterm, suffices.

The consequences of our findings are the following. 
Since three and higher-loop diagrams are superficially finite, these are
infrared dominated. Therefore, they are in principle calculable in the
classical theory. The loophole is of course the possible occurrence of
divergences in (one or two-loop) subdiagrams. To deal with these
divergences, counterterms have to be introduced. In the scalar case the
divergences occur only in the self-energy and are momentum independent,
therefore a mass renormalization is sufficient to obtain a cut-off
independent theory. This may be useful for a numerical approach 
to time-dependent problems, such as the dynamics of the phase transition
and/or topological defects in a (complex) scalar field theory.
In SU($N$) gauge theories the divergences are momentum
dependent, nevertheless a renormalization of the plasmon frequency
(\ref{coef}) takes care of the linear divergences. Two-loop divergences
cannot yet be handled, since we do not know what their precise form is. It
may be interesting to study them, not only for the introduction of
counterterms, but also to see if they have the same nice properties as the
one-loop divergences (classical HTL's), such as gauge invariance and a
conserved energy for the effective theory.

\subsubsection*{Acknowledgements}
It is a pleasure to thank Jan Smit for discussions. 
G.A.\ and B.N.\ thank the Institute for Nuclear Theory at the University
of Washington
for its hospitality and the Department of Energy for partial support
during the completion of this work.
G.A.\ was partly supported by FOM, the Netherlands, and by 
the TMR network {\em Finite Temperature Phase
Transitions in Particle Physics}, EU contract no.\ FMRX-CT97-0122.

\appendix

\renewcommand{\theequation}{\Alph{section}.\arabic{equation}}

\section{Hot, classical Feynman rules}
\setcounter{equation}{0}
\label{apprules}

\subsection{Scalar fields}
\label{appscalar}

In this appendix we discuss the construction of classical diagrams in
perturbation theory, i.e.\ the classical Feynman rules at finite
temperature, in scalar field theory. In order to do this, we start by
repeating some necessary ingredients of the approach that was introduced
in \cite{Aarts:1997kp}. 
For definiteness we use a massive scalar field with mass $m$, and 
interaction  
\be
 S_{\rm int} = -\int d^4x \left(\frac{1}{3!}g\phi^3 + 
\frac{1}{4!}\lm\phi^4\right).
\ee

In classical perturbation theory, two type of two-point functions appear. 
A perturbative solution of the equations of motion,
\be
\left(\partial_{t}^2-\nabla^2+m^2\right)\phi=
\frac{\delta S_{\rm int}}{\delta\phi},
\ee
 is constructed with 
the free retarded Green function $G_0^R(x)$, as
\be
\label{eqappsol}
 \phi(x) = \phi_0(x) + \int d^4x'\, G_0^R(x-x') 
\frac{\delta S_{\rm int}[\phi_0]}{\delta \phi_0(x')} 
+ \ldots,\ee
where $\phi_0(x)$ is the solution of the unperturbed problem with some  
arbitrary initial conditions. Note that this yields 
expressions in which only $\phi_0$'s are left over. 
The other two-point function specifies how to treat expectation values of
$\phi_0$'s.
The free thermal 
propagator $S_0(x)$ carries the thermal information and is defined by 
\be
S_0(x-x') = \bra\phi_0(x)\phi_0(x')\ket_{\rm cl,\, free}.
\ee
The brackets denote the averaging over the initial conditions weighted 
with the Boltzmann weight, for the unperturbed case \cite{Aarts:1997kp}.
In momentum space, the introduced two-point functions read
\bea
\label{eqgr}
&&\hspace{-0.7cm}G_0^R(K) = \frac{1}{\omk^2-(k^0+i\ep)^2} = \sum_{s=\pm} 
\frac{1}{2\omk}\frac{s}{k^0+i\ep + s\omk},\\
\label{eqS0}
&&\hspace{-0.7cm}S_0(K) = 
n_{\rm cl}(k^0)\ep(k^0)2\pi\delta(k_0^2-\omk^2) = 
\sum_{s=\pm} n_{\rm cl}(s\omk) \frac{1}{2\omk} 2\pi s \delta(k^0-s\omk),\\
&&\hspace{-0.7cm}n_{\rm cl}(k^0)=\frac{T}{\hbar k^0},
\;\;\;\;\omk=\sqrt{\veck^2+m^2},
\;\;\;\;\ep(k^0) = \theta(k^0)-\theta(-k^0).
\eea
The (free) retarded and (free) thermal two-point function are related by 
the classical KMS condition \cite{Parisi:1988,Aarts:1997kp}
\be
iS_0(K) = n_{\rm cl}(k^0)\left[G_0^R(K)-G_0^A(K)\right], 
\ee
where $G_0^A(K)=G_0^R(-K)$ is the free advanced Green function.
Finally, classical loop integrals containing these two-point functions 
arise from the spacetime integral(s) in (\ref{eqappsol}).

Explicitly solving the equations of motion perturbatively and making all
possible contractions to find all possible diagrams becomes rather
cumbersome at higher order in the coupling constants. Therefore we
discuss in the remainder of this appendix a set of rules which are based
on the underlying quantum perturbative approach. 

The imaginary-time or Matsubara formalism,
does not lead to a close connection 
with the classical approximation as described above at intermediate stages 
of a calculation.
However, a useful observation, obtained using the imaginary-time
formalism, is explained in section \ref{secdegree}. 
After performing the Matsubara sums, every diagram has a term
which has as many distribution functions as loops. 
Hence, in the classical limit $\hbar\to 0$ these terms remain. 
Other terms have less distribution functions and go to zero. This ensures us
that the classical limit is non-trivial and exists.

A formalism which lies closer to the classical perturbative approach is a
variation on the real-time formulation of finite temperature field theory,
and uses the closed time path (CTP) method
\cite{Schwinger:1961qe}. 
As is well-known, the CTP method involves a contour in the complex-time
plane
that consists of two branches, the upper branch ${\cal C}_+$ and the
lower branch ${\cal C}_-$ that runs back in time.  This leads to a
doubling of the fields, and they are denoted as $\phi_+,\phi_-$ to 
indicate on which branch they live. The propagator takes a matrix form, 
\be
{\bf G}(x-x') =
\left( \begin{array}{cc}
G^{++}(x-x') & G^{+-}(x-x')\\
G^{-+}(x-x') & G^{--}(x-x')   
\end{array}\right),
\ee
where the different superscripts specify the possible positions on 
and orderings along the contour.
A convenient variation is based on the Keldysh formalism, and is the
`center-of-mass/relative' coordinates version. It uses a 
change of basis from $\phi_{+,-}$ to $\phi_{1,2}$
\be
\label{eqrotphi}
 \left( \begin{array}{c} \phi_1\\
\phi_2
\end{array}\right) =
\left(\begin{array}{c}
(\phi_+ +\phi_-)/2
\\
\phi_+ - \phi_-
\end{array}\right) 
,
\ee
such that the (free) matrix propagator takes the form 
\cite{Aarts:1997kp} 
\be
\label{eqpropkel}
{\bf G}_{(0)}(x-x') \to 
\left( \begin{array}{cc} iF_{(0)}(x-x') & G^R_{(0)}(x-x')\\
G^A_{(0)}(x-x') & 0
\end{array}\right).
\ee
Here the free retarded and advanced Green function are given in momentum
space by the (classical) expression (\ref{eqgr}), and the quantum thermal
two-point function in momentum space reads
\be
F_0(K) = 
\sum_{s=\pm} [n(s\omk)+\half] \frac{1}{2\omk} 2\pi s 
\delta(k^0-s\omk),\;\;\;\; \nk=\frac{1}{\exp(\beta\hbar\omk)-1},
\ee
which is of course the quantum version of (\ref{eqS0}).
Again the (free) retarded and thermal two-point functions are 
related by the KMS condition 
\be
iF_0(K) = n(k^0)\left[G_0^R(K)-G_0^A(K)\right]. 
\ee

\begin{figure}
\centerline{\psfig{figure=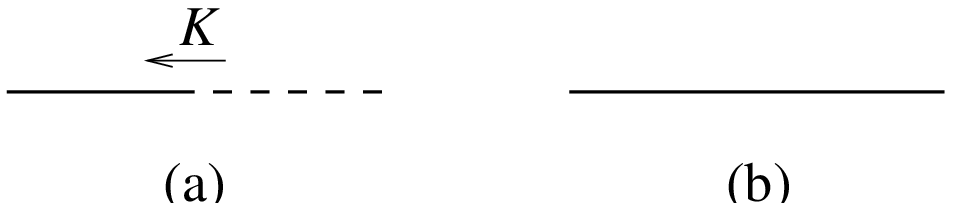,width=7cm}}
\caption{Propagators, (a)  $G_0^R(K) = G_0^A(-K)$, (b) $iF_0(K)$.}
\label{figprop}
\vspace{1cm}
\centerline{\psfig{figure=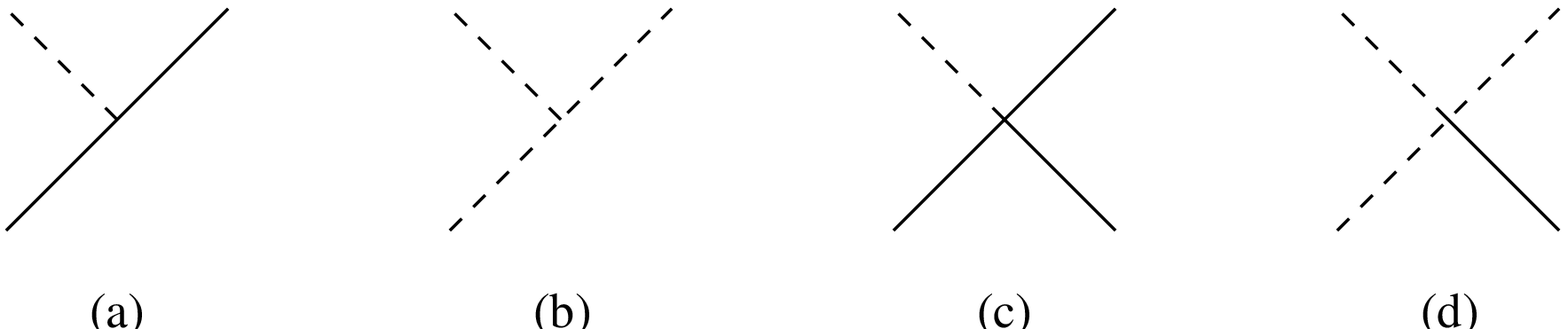,width=13cm}}
\caption{Vertices,
(a) $\frac{1}{2}g\phi_1^2\phi_2$, 
(b) $\frac{1}{4!}g\phi_2^3$, 
(c) $\frac{1}{3!}\lm\phi_1^3\phi_2$, 
(d) $\frac{1}{4!}\lm\phi_1\phi_2^3$.}
\label{figvertex} 
\end{figure}

Feynman rules appear when also the interaction part along the closed 
time path contour is written in terms of the $\phi_{1,2}$ fields 
\cite{VanEijck:1994}
\bea  
S_{\rm int} &=& -\int d^4x \left(
\frac{1}{3!}g\phi_+^3 - \frac{1}{3!}g\phi_-^3 +
\frac{1}{4!}\lm\phi_+^4 - \frac{1}{4!}\lm\phi_-^4  
\right) \nn\\
&=&
-\int d^4x \left(
\frac{1}{2}g\phi_1^2\phi_2 + \frac{1}{4!}g\phi_2^3 + 
\frac{1}{3!}\lm\phi_1^3\phi_2 + \frac{1}{4!}\lm\phi_1\phi_2^3  
\right).
\eea
The rules are presented pictorially in figs.\ \ref{figprop} and 
\ref{figvertex}. The $\phi_1$ field is denoted with a full line and 
the $\phi_2$ field with a dashed line. For the retarded and advanced Green 
functions, it is necessary to specify the direction of the momentum flow 
through the propagator, and this is indicated with the arrow.

\begin{figure}
\centerline{\psfig{figure=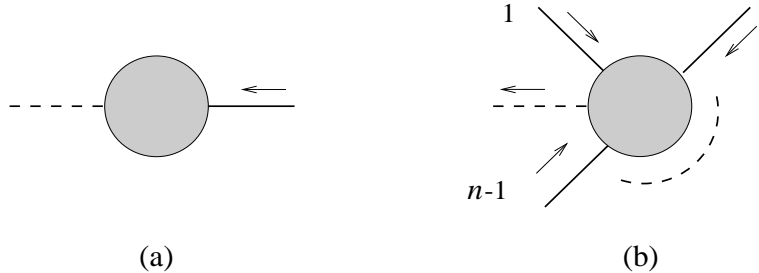,width=10cm}}
\caption{(a) Retarded self-energy, (b) generalized retarded $n$-point
vertex functions.}
\label{figbubbles}
\end{figure}

The retarded self energy and the so-called generalized retarded $n$-point 
vertex functions \cite{VanEijck:1994} have one dashed `leg' and $n-1$ full
`legs'. 
These are shown in fig.\ \ref{figbubbles}. The arrows denote again 
the momentum flow of the external momenta. 

We now discuss the \(\hbar\rightarrow 0\) limit of these real-time quantum 
Feynman rules.
This limit affects the diagrams in two ways. The first one is
obvious, the thermal propagator $F_0$ has to be replaced by $S_0$. The
second change leads to a drastic simplification: only the vertices (a) and
(c) in fig.\ \ref{figvertex} contribute in the classical limit, and the
two other vertices (b) and (d) can be neglected. This can be seen as
follows: vertices (b) and (d) can only appear in a diagram with retarded
(or advanced) Green functions attached to the three dashed legs. 
After attaching these Green functions, the resulting outer lines (which 
either still have to be attached to another vertex or are external lines) 
are always full lines. 
However, such a configuration can be constructed as well with vertices (a)
and (c): these vertices have two full legs where (b) resp.\ (d) have two
dashed legs. By attaching two thermal two-point functions on these legs, the
external lines are full as well, and the vertices can be part of a 
diagram in exactly the same manner. But a classical thermal two-point 
function is proportional to $1/\hbar$. Diagrams with vertex (a) or 
(c) have two more thermal two-point functions than the corresponding 
diagrams with vertex (b) or (d). Hence, the first class of diagrams is 
relatively stronger in the classical limit with respect to the second 
class by a factor $1/\hbar^2$.\footnote{Negative powers of $\hbar$ will of 
course be canceled by positive powers coming from loop counting.} In 
other words, vertices (b) and (d) will be ${\cal O}(\hbar^2)$ suppressed 
with respect to vertices (a) and (c).

We propose that classical Feynman rules follow from the quantum ones by taking 
\(\hbar\) to zero, which results in the following (simple) rules:
\begin{enumerate}
\item Draw all diagrams as in the quantum case, but use only vertices (a) 
and (c).
\item Replace the thermal propagator $F_0$ by its classical counterpart 
$S_0$.
\end{enumerate}
An explicit check of these rules (by a comparison with the results 
obtained by perturbatively solving 
the equations of motion and averaging over the initial conditions) can be 
found for
the case of $\lm\phi^4$-theory for the two-point function up to two loops 
and the four-point function to one loop in \cite{Aarts:1997kp}.

\subsection{Gauge invariant cut-off in the classical theory}

\label{appgaugecutoff}

We argue that in classical gauge theories it is possible to  introduce a 
(continuum) momentum cut-off without breaking gauge invariance. 
The basic ingredient is the result of Landshoff and Rebhan 
\cite{Landshoff:1992ne} that in general linear
gauges it is possible to formulate a (quantum) 
real-time theory in which only the two physical degrees of freedom 
of the gauge field acquire a thermal part.
This implies that a change in the distribution function 
\be
n(k^0)\rightarrow n(k^0)f(k/\Lambda),
\label{cutoffdistr}
\ee
with $f$ some function, does not break gauge invariance. Introducing a
cut-off in this way will not affect the Slavnov-Taylor identities.
This has been employed in a Wilson renormalization group approach
to hot (quantum) SU($N$) gauge theories
\cite{D'Attanasio:1996fy}.  

If we take the classical limit of (\ref{cutoffdistr}) and choose \(f\) as
the step function, we get
\be
n_{\rm cl}(k^0)\rightarrow n_{\rm cl}(k^0)\theta(\Lambda-k),
\label{clcutoff}
\ee
which as (\ref{cutoffdistr}) does not break gauge invariance. It is for
instance straightforward to check that the HTL's calculated with
distribution function (\ref{clcutoff}) satisfy the same abelian-like Ward
identities as usual. Finally we should remark that the regularization
(\ref{clcutoff}) is sufficient to render the theory ultraviolet finite,
since each loop introduces one distribution function.\footnote{In the
quantum theory the cut-off in (\ref{cutoffdistr}) acts only on thermal
fluctuations. A zero-temperature regularization and renormalization is
still necessary to avoid divergences coming from the zero-temperature
quantum fluctuations.}

\section{Classical one-loop SU($N$) self-energy: explicit calculation}
\setcounter{equation}{0}
\label{appcalculation}

We present in this appendix the calculation of the classical self-energy
in SU($N$) gauge theory, in particular the $\Pi_{ii}$ part, in the Feynman
gauge. The starting point is given by (\ref{onels-e}) in the main text.
After changing variables from $\veck \to -\veck-\vecp$ in the part that is 
proportional to $\npkcl$, we find
\be
\Pi_{ii,\rm cl}^{ab}(P) = \delta^{ab}g^2N\Pi_{\rm cl}(P),
\ee
with
\be
 \Pi_{\rm cl}(P) = \intk\, \frac{\hbar\nkcl}{\omk}
\left\{6+
\frac{A_{ii}}{(p^0+\omk)^2-\ompk^2} + \frac{A_{ii}}{(p^0-\omk)^2-\ompk^2}
\right\},
\ee
and $A_{ii} = 4k^2+4\veck\cdot\vecp+5p^2-6p_0^2$. We have combined 
$\hbar$ with $\nkcl$, which is an $\hbar$-independent combination.

The angular integrations can be performed, and 
\bea
\label{eqself}
&&\Pi_{\rm cl}(p^0,p) = 
\int dk\, \frac{k\hbar\ncl}{\pi^2}\bigg\{ 
1+\frac{p^0}{p}\ln \frac{p_+}{p_-} 
- \frac{k}{2p}\Big[L_+(k)-L_-(k)\Big]\bigg\}\\ 
&&\;\;\;\;\;\;\;\;-
\nonumber
\int dk\, \frac{k\hbar\ncl}{8\pi^2p}\bigg\{
\frac{3p^2-4p_0^2}{k}\Big[L_+(k)-L_-(k)\Big] + 
4p^0\Big[L_+(k)+L_-(k)\Big]\bigg\}. 
\eea
Motivated by Weldon \cite{Weldon:1982aq}, we used here the notation
\be 
p_\pm=\half(p^0\pm p),\;\;\;\; L_\pm(k) = \ln \frac{k\pm p_+}{k\pm p_-}.
\ee
The result (\ref{eqself}) agrees with the expression obtained by Weldon 
in the appendix of \cite{Weldon:1982aq}, except of course that the 
distribution function is classical in our case.

The remaining radial integral in the first line of (\ref{eqself}) is 
linearly 
divergent. For the first two terms this is obvious, and for the third 
term one can use $L_+(k)-L_-(k) = 2p/k +{\cal O}(k^{-3})$. In fact, the 
divergence in this term cancels against the first term.
The integrals in the second line are convergent. 
To regulate the divergences, we use the distribution function with 
a momentum cut-off $\hbar\ncl=T/k\;\theta(\Lambda-k)$. The final result 
requires the evaluation of four integrals, which read (recall that $p^0$ 
contains a small positive imaginary part) 
\bea
&&\int dk\, k\hbar\ncl = T\Lambda,\\
&&\int dk\, k^2\hbar\ncl 
\Big[L_+(k)-L_-(k)\Big] = T\left(2p\Lambda+\half \pi i pp^0\right),\\
&&\int dk\, \hbar\ncl \Big[L_+(k)-L_-(k)\Big] = T\pi i 
\ln\frac{p_+}{p_-},\\
&&\int dk\, k\hbar\ncl \Big[L_+(k)+L_-(k)\Big] = -T\pi ip.
\eea
The second and fourth integral are straightforward using partial 
integration, and the third one can be performed by complex contour 
integration while being careful around $k=0$. Note that these integrals 
are much simpler than in the quantum case, because of the simple $k$ 
dependence of the classical distribution function.

Putting all the results together, we find for the classical one-loop 
retarded self energy
\be
\Pi_{ii,\rm cl}^{ab}(P) =
N\delta^{ab}g^2\left[ \frac{T\Lambda}{\pi^2}\frac{p^0}{p}\ln 
\frac{p_+}{p_-} +
\frac{T}{4\pi}\bigg( ip^0-\frac{3p^2-4p_0^2}{2p}i\ln 
\frac{p_+}{p_-}\bigg)\right],
\ee
which is presented in (\ref{eqSUn}).

\section{Two loop naively linear divergent contributions}
\setcounter{equation}{0}
\label{appnaive}
\subsection{Diagram b}
In this appendix we give the results for the naively linearly divergent
contributions to the classical two-loop self-energy. We start with the
classical limit of the self-energy diagram (b) in fig.\
\ref{figtwoloopall}, presented in (\ref{twoloopex2}), and use the
shorthand notation of (\ref{eqshorthand}). There are three naively
linearly divergent contributions and we shall denote these with (b1),
(b2), and (b3).

We start with contribution (b1), obtained by taking 
$s_3=s', s_2=-s$ and setting the external $p^0, \vecp$ to zero in the
energy denominators with three loop-energies. We then find
\bea
&&
\Pi^{\rm (b1)}_{ij, \rm cl}(P) =
\half (g^2\hbar)^2 \intk\intkp (k)^4_{ij} \sum_{ss's_1}
\frac{-s_1}{2^5\om\om'\om_1\om_2\om_3}\,
\frac{1}{p^0+s'(\om'+\om_3)}
\nn\\
&&\hspace{0.5cm}
\frac{n_{\rm cl}(s\om_2)-n_{\rm cl}(s\om)}{p^0+s(\om-\om_2)}\,
\left[
\frac{n_{\rm cl}(s_1 \om_1) - n_{\rm cl}(s'\om_3)}{-s'\om'-s\om+s_1\om_1}
-\frac{n_{\rm cl}(s_1 \om_1) + n_{\rm cl}(s'\om')}{s'\om'-s\om+s_1\om_1}
\right].
\label{IA}
\eea
The difference between distribution functions 
\(\left[n_{\rm cl}(s\om_{\vecp-\veck})-n_{\rm cl}(s\omk)\right]\) 
reduces the degree of divergence by one compared to the naive
estimate, which is from linear to logarithmic.
Note that the other difference between distribution functions
\(\left[n_{\rm cl}(s_1 \om_{\veck-\veck'}) -
n_{\rm cl}(s'\om_{\vecp-\veck'})\right]\),
does not reduce the degree of divergence any further, since \(\veck\)
is not a (small) external momentum, but is integrated over.

A similar contribution is obtained by taking \(s_2=s\) and \(s_3=-s'\)
and again setting \(p^0,\vecp=0\) in the same energy denominators. We 
obtain
\bea
&&\Pi^{\rm (b2)}_{ij, \rm cl}(P) = 
\half (g^2\hbar)^2\intk\intkp \, (k)^4_{ij}
\sum_{s s^{\prime} s_{1}}
\frac{-s_1}{2^5\om\om'\om_1\om_2\om_3}\,
\frac{1}{p^0+s'(\om'-\om_3)}
\nn\\
&&\hspace{0.5cm}
\frac{n_{\rm cl}(s'\om_3) - n_{\rm cl}(s'\om')}{p^0+s(\om+\om_2)}
\left[
-\frac{n_{\rm cl}(s_1 \om_1)+n_{\rm cl}(s\om_2)}{s'\om'+s\om+s_1\om_1}
+\frac{n_{\rm cl}(s_1\om_1)-n_{\rm cl}(s\om)}{s'\om'-s\om+s_1\om_1}
\right].
\label{IB}
\eea
Again a difference between distribution functions appears that reduces the 
degree of divergence to a logarithmic one.

The third naively linearly divergent contribution to consider is of a
different type. It is obtained from the classical limit of
(\ref{twoloopex2})  by setting \(s=-s_2\) and \(s'=-s_3\) and taking the
linear term in \(p^0,\vecp\) in an expansion of the energy denominator
with \(\om_1 = \om_{\veck-\veck'}\). The zeroth order term in this
expansion gives rise to a naively quadratic divergence and was already
discussed in the main text. The first-order term reads
\bea
\Pi^{\rm (b3)}_{ij, \rm cl}(P) 
\!\!\!\!&=&\!\!\!\!
 \half (g^2\hbar)^2 \intk\intkp\, (k)^4_{ij} \sum_{ss's_{1}}  
\frac{-s_1}{2^5\om\om'\om_1\om_2\om_3}\,
\nn\\
&&\!\!\!\!
\frac{1}{p^0+s'(\om'-\om_3)}\,
\frac{1}{p^0+s(\om-\om_2)}\,
\frac{1}{\left(s'\om'-s\om+s_1\om_1\right)^2}
\nn\\
&&\!\!\!\!
\bigg\{
s'(\vecp\cdot\khat')
\left[n_{\rm cl}(s\om_2)-n_{\rm cl}(s\om)\right]
\left[n_{\rm cl}(s'\om_3) + n_{\rm cl}(s_1\om_1)\right]\nn\\
&&
-s(\vecp\cdot\khat)
\left[n_{\rm cl}(s'\om_3) - n_{\rm cl}(s'\om')\right]
\left[n_{\rm cl}(s\om_2) -n_{\rm cl}(s_1\om_1)\right]
\nn\\
&&
+p^0
n_{\rm cl}(s_1\om_1)
\Big([n_{\rm cl}(s\om)-n_{\rm cl}(s\om_2)] +
[n_{\rm cl}(s'\om')-n_{\rm cl}(s'\om_3)]\Big)
\nn\\
&&
+p^0
n_{\rm cl}(s_1\om_1)
\left[n_{\rm cl}(s\om)n_{\rm cl}(s'\om_3) - 
n_{\rm cl}(s\om_2)n_{\rm cl}(s'\om')\right]
\bigg\}.
\label{IC}
\eea
We emphasize again that the region of phase space where 
$s'\om'-s\om+s_1\om_1$ vanishes is excluded in this expansion. 
The first three terms between curly brackets all have a
factor which is the difference between distribution functions. The fourth
term is different, but also here the factor with distribution functions  
vanishes when the external momentum is taken to zero (i.e.\
when $\om_2\to\om,
\om_3\to\om'$).
Hence this factor contributes a power $\Lambda^{-3}$ instead of
$\Lambda^{-2}$, and it brings down the degree of divergence.
We conclude that the degree of divergence is reduced from linear to
logarithmic in contribution (b3) as well.

\subsection{Diagram c}
The final diagram that needs to be examined is diagram (c) in fig.\
\ref{figtwoloopall}. The quantum expression is
\bea
&&
\Pi^{\rm (c)}_{ij}(P) = (g^2\hbar)^2 \intk\intkp\, (k)^2_{ij}
\sum_{s s' s_1 s_2}
\frac{ss's_1 s_2}{2^4 \om\om'\om_1\om_2}\,
\frac{1}{p^0-s\om-s_2\om_2}\nn\\
&&
\hspace{2.4cm}
\bigg\{
\frac{1}{s\om+s'\om'-s_1\om_1}
\Big([n(s\om)+1][n(s'\om')+1]n(s_1\om_1)
\nn\\
&&\hspace{3.8cm}
- n(s\om)n(s'\om')[n(s_1\om_1)+1]\Big)\nn\\
&&
\hspace{2.4cm}
+
\frac{1}{p^0-s_2\om_2+s'\om'-s_1\om_1}
\Big( [n(s'\om')+1]n(s_1\om_1)n(s_2\om_2)
-
\nn\\
&&
\hspace{3.8cm}
n(s'\om')[n(s_1\om_1)+1][n(s_2\om_2)+1]\Big)
\bigg\},
\label{diagramc}
\eea
where in this case $\om_1=\om_{\vecp-\veck-\veckp}$ and we inserted
\((k)^2_{ij}\) to indicate the two powers of momentum that
come from the two three-point vertices.

We take the classical limit of (\ref{diagramc}). 
The contribution with \(s_2=-s\) 
is naively linearly divergent, it reads
\bea
&&
\tilde{\Pi}^{\rm (c)}_{ij,\rm cl} = (g^2\hbar)^2 \intk\intkp\,
(k)^2_{ij}
\sum_{s s' s_1 }
\frac{-s's_1 }{2^4 \om\om'\om_1\om_2}\,
\frac{1}{p^0-s\om+s\om'}
\nn\\
&&\hspace{1cm}
\frac{1}{s\om+s'\om'-s_1\om_{\veck+\veckp}}\,
[n(s\om)-n(s\om_2)][n(s_1\om_1)-n(s'\om')].
\eea
Again the first difference between distribution functions reduces the
degree of divergence to a logarithmic one.

\providecommand{\href}[2]{#2}\begingroup\raggedright\endgroup


\end{document}